\documentclass[11pt]{article}
\usepackage{Moriond,epsfig}

\def\Journal#1#2#3#4{{#1} {\bf #2}, #3 (#4)}

\def\P2{\em arXiv:0710.3050v1}

\def\NPB{{\em Nucl. Phys.} B}
\def\PL{\em Phys. Lett.}
\def\PLB{{\em Phys. Lett.}  B}
\def\PRL{\em Phys. Rev. Lett.}
\def\PRD{{\em Phys. Rev.} D}

\def\EPJC{{\em Eur. Phys. J.} C}
\def\PR{\em Phys. Rep.}
\def\JHEP{\em JHEP}

\def\PPNL{\em Phys. Part. Nucl. Lett.}
\def\Proc{\em arXiv:0803.4475v1}
\def\Gas1{\em arXiv:0710.3048v1}

\def\be{\begin{equation}}
\def\ee{\end{equation}}
\def\bea{\begin{eqnarray}}
\def\eea{\end{eqnarray}}

\begin{document}
\vspace*{4cm}
\title{NEW RESULTS ON KAON DECAYS FROM NA48/2}

\author{CRISTINA MORALES MORALES~\footnote{On behalf of the NA48/2 Collaboration.}}

\address{Institut f\"ur Physik, Johannes Gutenberg-Universit\"at, 55099 Mainz, Germany\\
E-mail: cmorales@mail.cern.ch}

\maketitle\abstracts{Recent results from the NA48/2 experiment are
presented. The $\pi\pi$ scattering lengths $a_0^0$ and $a_0^2$ have
been extracted from the cusp in the $M_{00}^2$ distribution of
$K^\pm \rightarrow \pi^\pm \pi^0 \pi^0$ decays and from the $K^\pm
\rightarrow \pi^+ \pi^- e^\pm \nu$ phase shift $\delta$. Branching
ratios and form factors have been measured for \mbox{$K^\pm
\rightarrow \pi^\pm \gamma \gamma$}, \mbox{$K^\pm \rightarrow
\pi^\pm \gamma~e^+ e^-$} and \mbox{$K^\pm \rightarrow \pi^\pm e^+
e^-$} decays and are also summarized here.}

\section{Introduction}
During 2003 and 2004, the NA48/2 experiment at CERN SPS has
collected the world largest amount of charged kaon decays. The main
goal of NA48$/$2 was the search for direct CP violation in $K^{\pm}$
decays into three pions. However, given the high statistics
achieved, many other physics topics were also covered including the
study of the $\pi\pi$ interaction at low energy, radiative decays,
the measurement of $V_{us}$ from semileptonic decays, etc.. In the
following sections, recent results on ChPT parameters obtained by
the NA48/2 Collaboration will be presented.

\section{The NA48/2 experiment}
Simultaneous $K^+$ and $K^-$ beams were produced by 400 GeV protons
from the CERN SPS, impinging on a Be target. Kaons were deflected in
a front-end achromat to select a momentum band of 60 $\pm$ 3~GeV/$c$
and then focused such that they converge about 200 m downstream at
the beginning of the detector. A description of the detector can be
found in \cite{det}. For the measurements presented here, the most
important detector components are the magnet spectrometer,
consisting of two drift chambers before and two after a dipole
magnet, and the quasi-homogeneous liquid krypton calorimeter. The
momentum of the charged particles and the energy of the photons are
measured with a relative uncertainty of $1\%$ at 20 GeV. The trigger
was mainly designed to select events with three charged tracks
(charged trigger) and $K^\pm \rightarrow \pi^\pm \pi^0 \pi^0$ events
(neutral trigger).

\begin{figure}[t]
  \begin{minipage}[b]{7 cm}
\epsfig{figure=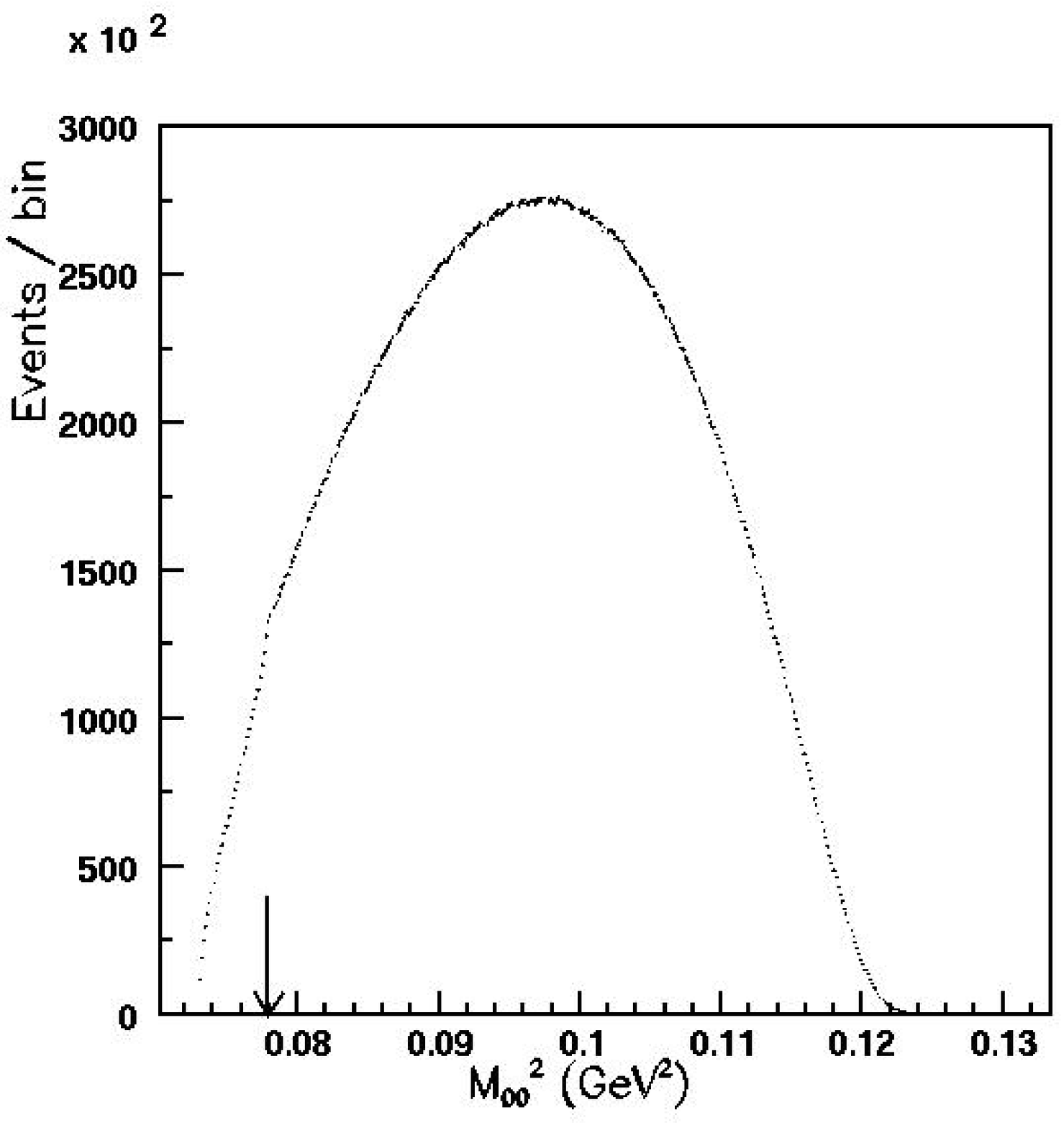,width=7cm,height=5.5cm}
  \end{minipage}
  \begin{minipage}[b]{7 cm}\hspace{0.5cm}
  \epsfig{figure=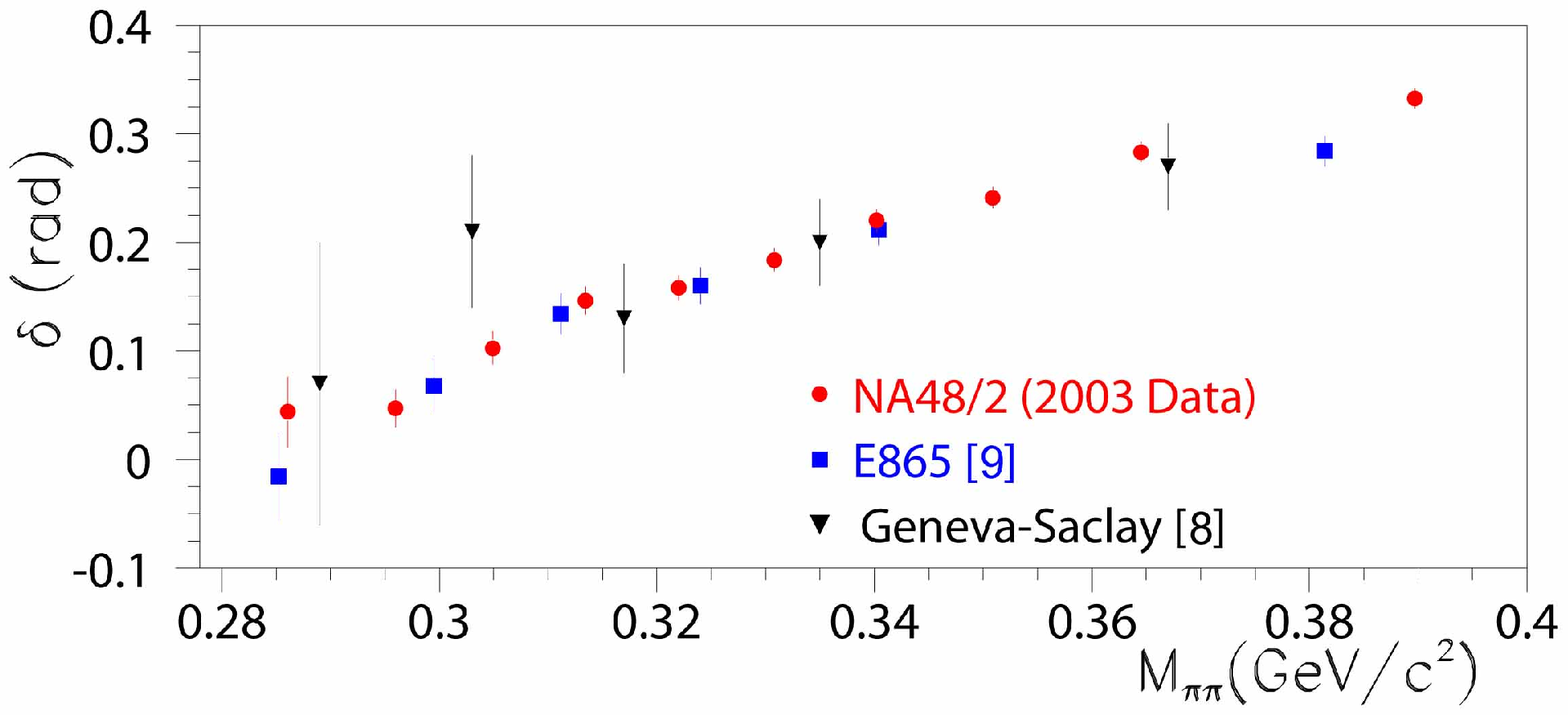,width=8.5cm,height=5.5cm}
  \end{minipage}
  \caption[]{(Left) Invariant $\pi^0 \pi^0$ mass squared of $K^\pm \rightarrow \pi^\pm \pi^0 \pi^0$
  candidates. Note the presence of a cusp for $M^2_{00} = 4m^2_{\pi^+}$ (arrow).
  (Right) Variation of phase shift in $K^\pm \rightarrow \pi^+ \pi^- e^\pm \nu$ decays with $\pi^+\pi^-$ invariant mass. \label{fig:pipiscattering}}
\end{figure}

\section{Measurement of $\pi\pi$ scattering lengths}

The quark condensate $\langle 0 | \overline{q}q | 0 \rangle$ is a
fundamental parameter of ChPT. Its value must be determined
experimentally, e.g. by measuring the $\pi\pi$ scattering lengths
$a_0^0$ and $a_0^2$,
which are predicted very precisely within the framework of ChPT~\cite{Co}.\\
NA48/2 has reported two new measurements of the $\pi\pi$ scattering
lengths using \mbox{$K^\pm \rightarrow \pi^\pm \pi^0 \pi^0$} and
\mbox{$K^\pm \rightarrow \pi^+ \pi^- e^\pm \nu$} decays. A cusp
observed in the $M_{\pi^0\pi^0}$ distribution of
\mbox{{\boldmath{$K^\pm \rightarrow \pi^\pm \pi^0 \pi^0$}}} \rm
decays at $M^2_{00} = 4m^2_{\pi^\pm}$ (Fig.~\ref{fig:pipiscattering}
(left)) can be explained by $\pi^+\pi^-$ re-scattering
terms~\cite{Meis,Cab} and provides a measurement of $a_0^0$ and
$a^2_0$ from a fit of the $M^2_{00}$ distribution around the cusp
discontinuity. A sample of about $59.6 \times 10^6$ decays from 2003
and 2004 data has been used for this analysis, and the preliminary
results from the fit of the Cabibbo-Isidori model~\cite{Cab2} are:
\begin{equation}\nonumber
\begin{array}{rcl}
(a^0_0 - a^2_0)m_{\pi^+} & = & 0.261 \pm 0.006_{stat} \pm
0.003_{syst} \pm 0.001_{ext} \pm 0.013_{theory},\\
\nonumber a_0^2m_{\pi^+} & = & -0.037 \pm 0.013_{stat} \pm
0.009_{syst} \pm 0.002_{ext}, \nonumber
\end{array}
\end{equation}
where the theoretical uncertainty is due to neglected $O(a_i^3)$ and
radiative corrections. Alternative fits are being performed
following the approach by \cite{Col}.\\
In {\boldmath{$K^\pm \rightarrow \pi^+ \pi^- e^\pm \nu$}} \rm
decays, the pions are produced close to threshold. The decay
amplitude depends on the complex phases $\delta_0$ and $\delta_1$
(the $S$ and $P$ waves $\pi\pi$ phase shifts for isospin $I = 0$).
The difference $\delta = \delta_0 -\delta_1$ can be measured as a
function of the invariant mass of the two pions, $M_{\pi\pi}$.
NA48/2 has performed a combined fit to the decay form factors and
the phase shift difference as a function of $M_{\pi\pi}$ in a sample
of 670000 signal candidates with $0.5\%$ background~\cite{B}. The
results are shown in Fig.~\ref{fig:pipiscattering} (right) together
with two earlier experiments~\cite{6,7}. From the phase shift
measurements, the $\pi\pi$ scattering lengths can be extracted using
dispersion relations~\cite{R}. At the center of the Universal
Band~\cite{UB}, $a_0^2$ is related to $a_0^0$. A one paramater fit
gives $a_0^0 = 0.256 \pm 0.006_{stat} \pm 0.002_{syst}
{}^{+0.018}_{-0.017}{}_{ext}$, which implies $a_0^2 = -0.0312 \pm
0.0011_{stat} \pm 0.0004_{syst} {}^{+0.0129}_{-0.0122}{}_{ext}$. The
external error reflects the width of the Universal Band. From a two
parameters fit, the results are:
\begin{equation}\nonumber
\begin{array}{rcl}
a^0_0m_{\pi^+} & = & 0.233 \pm 0.016_{stat} \pm 0.007_{syst}, \\
\nonumber a_0^2m_{\pi^+} & = & -0.047 \pm 0.011_{stat} \pm
0.004_{syst},
\end{array}
\end{equation}
with $\rho = 0.967$. Theoretical work including isospin symmetry
breaking effects~\cite{G1}suggests that $a_0^0$ could decrease by
$\approx$ 0.02 for and $a_0^2$ by $\approx$ 0.004, bringing this
measurement in agreement with other measurements and ChPT
predictions~\cite{B}.


\begin{figure}[t]
  \begin{minipage}[b]{7 cm}
  \epsfig{figure=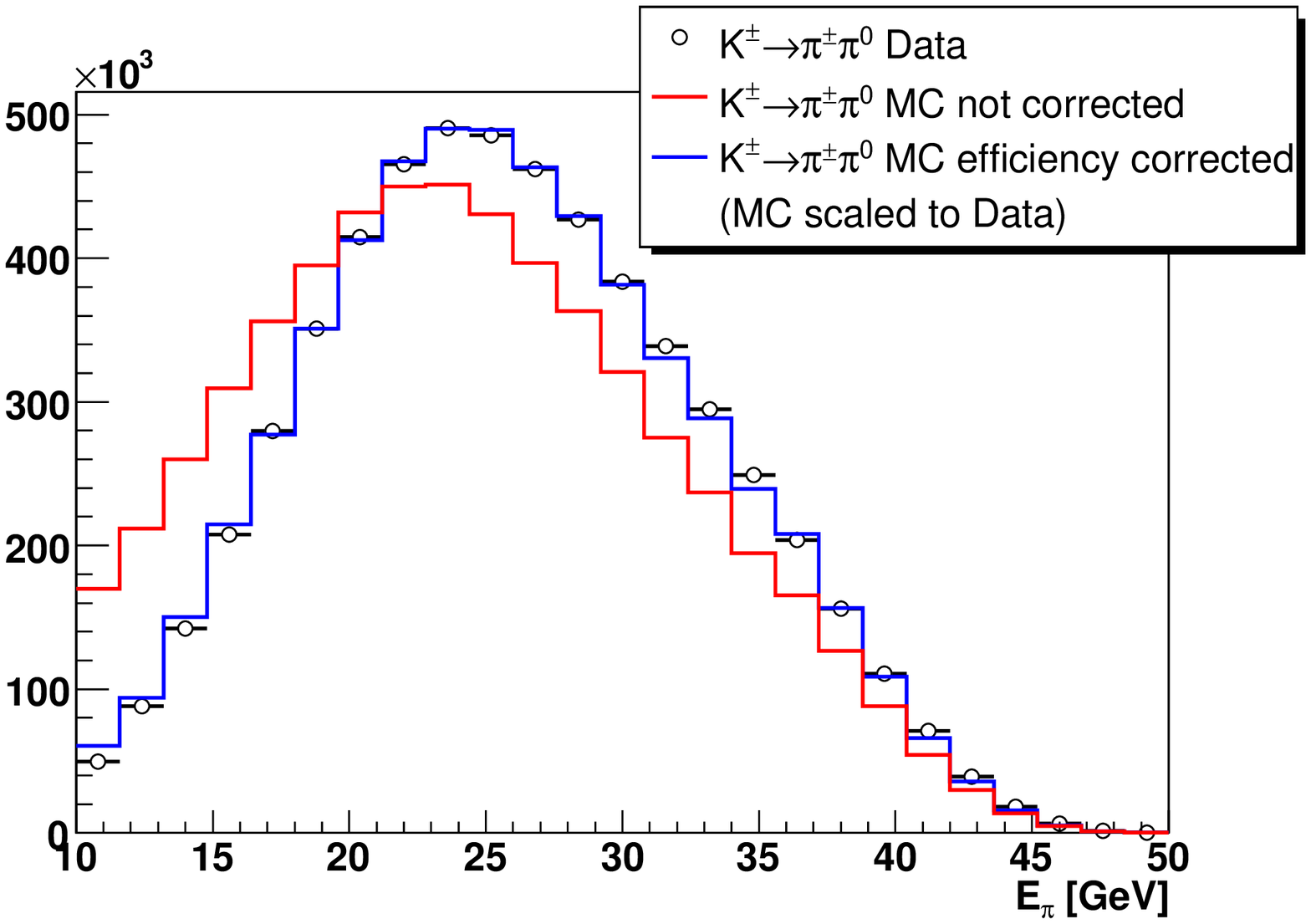,width=8.25cm,height=5.5cm}
  \end{minipage}
  \begin{minipage}[b]{7 cm}\hspace{1cm}
  \epsfig{figure=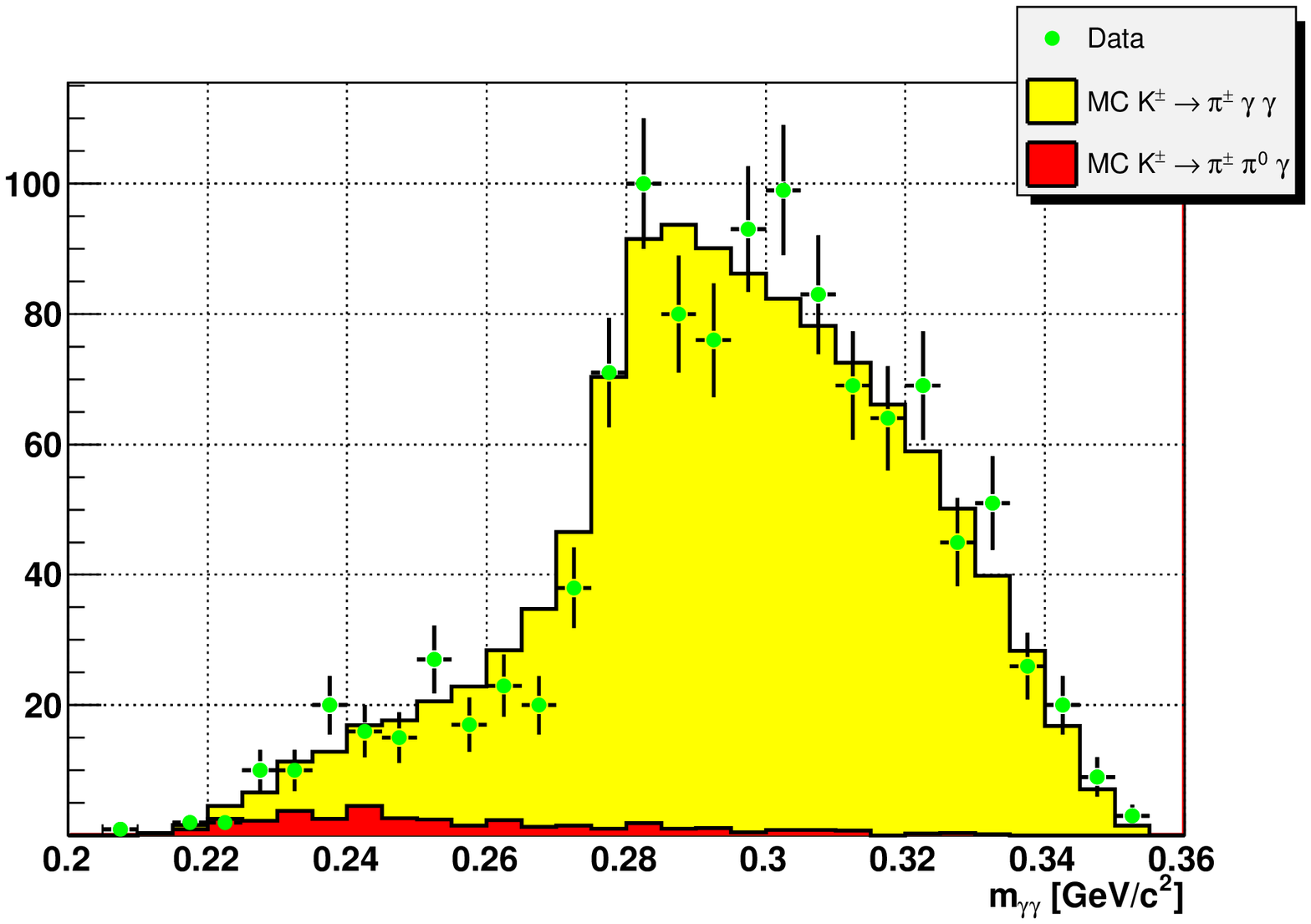,width=8.3cm,height=5.5cm}
  \end{minipage}
  \caption[]{(Left) Pion track energy of $K^\pm \rightarrow \pi^\pm
  \pi^0$ normalization data (black) and MC events (red, blue) without and with trigger
  efficiency correction, respectively.
  (Right) $M_{\gamma \gamma}$ invariant mass of $K^\pm \rightarrow \pi^\pm \gamma \gamma$ candidates.\label{fig:pigg}}
\end{figure}

\section{$K^\pm \rightarrow \pi^\pm \gamma \gamma$ analysis}
The contributions of the chiral lagrangian to this decay~\cite{Pich}
appear at $O(p^4)$. At this order, only the $\Delta I = 1/2$
invariant amplitudes $A(z)$ and $C(z)$ with $z =
M^2_{\gamma\gamma}/M^2_{K^\pm}$ contribute. $A(z)$ contains the
\emph{$O(p^4)$ loop diagram} contributions and the \emph{tree level
counterterms} absorbed in unknown parameter $\hat{c}$ predicted to
be positive and of $O(1)$~\cite{G2}. The loop leads to a
characteristic signature in the invariant mass $M_{\gamma \gamma}$
distribution, which is favoured to be above $2m_{\pi^+}$ and
exhibits a cusp at $2m_{\pi^+}$ threshold. The parameter $\hat{c}$
fixes the value of the branching ratio and the $M_{\gamma\gamma}$
spectrum shape. $C(z)$ contains \emph{poles and
tadpoles}~\cite{Pich,Trine} effects. $O(p^6)$ studies
concluded~\cite{Jorge} that unitarity correction effects could
increase the BR between $30\% - 40\%$, while vector
meson exchange contributions would be negligible.\\
NA48/2 has analyzed about $40\%$ of its data, finding 1164 signal
candidates with $3.3\%$ background (40 times more statistics than
previous experiments~\cite{BNL}). This decay and its normalization
channel ($K^\pm \rightarrow \pi^\pm \pi^0$) were collected through
the neutral trigger chain intended for the collection of
\mbox{$K^\pm \rightarrow \pi^\pm \pi^0 \pi^0$} decays and therefore
suffered from a very low trigger efficiency ($\approx 50\%$).
Elaborate studies were performed to measure these efficiencies and
correct for them (see Fig.~\ref{fig:pigg} (left)). The reconstructed
$M_{\gamma \gamma}$ spectrum can be seen in Fig.~\ref{fig:pigg} for
selected candidates (crosses), signal MC (yellow) and background (red).\\
The model dependent branching ratio of \mbox{$K^\pm \rightarrow
\pi^\pm \gamma \gamma$} has been measured, assuming the validity of
the $O(p^6)$ ChPT as presented in~\cite{Jorge} and taking $\hat{c} =
2$~\footnote[1]{This is a realistic assumption based on previous
results by~\cite{BNL} which obtained $\hat{c}$ = 1.8 $\pm$ 0.6.}.
The preliminary result is $BR(K^\pm \rightarrow \pi^\pm \gamma
\gamma) = (1.07 \pm 0.04_{stat} \pm 0.08_{syst})\times 10^{-6}$. A
model independent BR measurement is in preparation, together with
the extraction of $\hat{c}$ from a fit to $M_{\gamma \gamma}$ and
BR.

\section{$K^\pm \rightarrow \pi^\pm \gamma~e^+ e^-$ analysis}

\begin{figure}[t]
  \begin{minipage}[b]{8 cm}
  \epsfig{figure=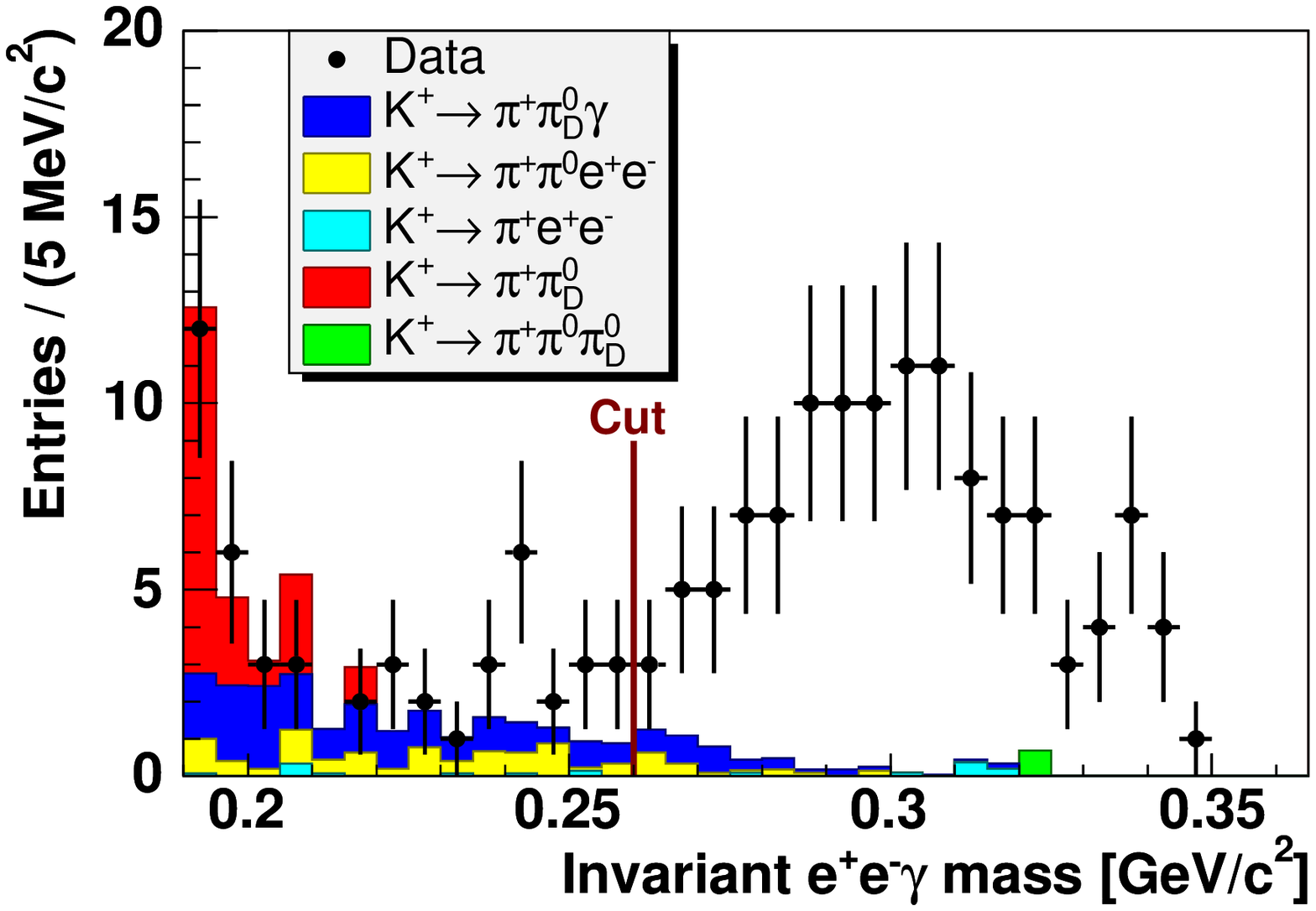,width=9cm,height=5.5cm}
  \end{minipage}
  \begin{minipage}[b]{7 cm}\hspace{1cm}
  \epsfig{figure=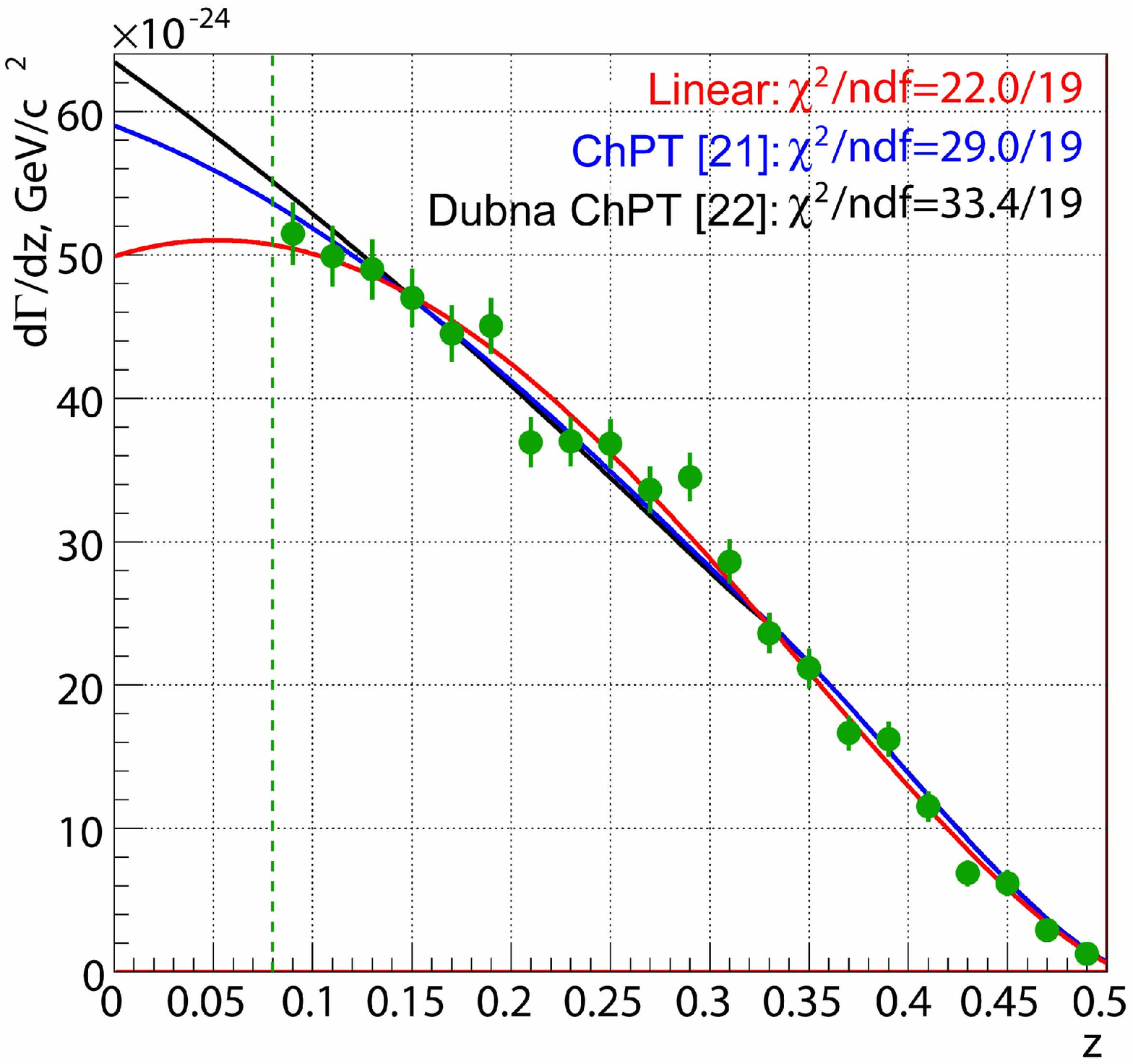
  ,width=6.5cm,height=5.5cm}
  \end{minipage}
  \caption[]{(Left) $M_{\gamma~e^+ e^-}$ invariant mass of $K^\pm \rightarrow
\pi^\pm \gamma~e^+ e^-$ candidates. Crosses are signal and colored
histograms background. (Right) $K^\pm \rightarrow \pi^\pm e^+ e^-$
differential decay rate and different
    fit results from the considered models. \label{fig:pieeg}}
\end{figure}

This decay is similar to $K^\pm \rightarrow \pi^\pm \gamma \gamma$
with one photon internally converting into a pair of electrons.
NA48/2 has reported the first observation of the decay \mbox{$K^\pm
\rightarrow \pi^\pm \gamma~e^+ e^-$} using the full 2003 and 2004
data sample~\cite{Ca}. 120 candidates with $7.3 \pm 1.7$ estimated
background events have been selected in the accessible region with
$M_{\gamma e e} > 0.26$~GeV/$c^2$ invariant mass. The candidates are
shown in Fig.~\ref{fig:pieeg} (left). Using $K^\pm \rightarrow
\pi^\pm \pi^0_D$ as normalization channel, the branching ratio has
been determined in a model independent way to be $\mathrm{BR} =
(1.19 \pm 0.12_{stat} \pm 0.04_{syst}) \times 10^{-8}$ for
$M_{\gamma e e } > 0.26~\mathrm{GeV}/c^2$. The parameter $\hat{c}$
has also been measured assuming the validity of $O(p^6)$~\cite{Ga}
and found to be $\hat{c} = 0.90 \pm 0.45$.


\section{$K^\pm \rightarrow \pi^\pm e^+ e^-$ analysis}
The FCNC process $K^\pm \rightarrow \pi^\pm e^+ e^-$
can be described in ChPT~\cite{M1}. NA48/2 has collected 7146
candidates with $0.6\%$ background. The decay rate has been measured
using \mbox{$K^\pm \rightarrow \pi^\pm \pi^0_D$} as normalization. A
preliminary model independent measurement for $z =
M^2_{e^+e^-}/M^2_{K^\pm} > 0.08$ gave $BR = (2.26 \pm 0.03_{stat}
\pm 0.03_{syst} \pm 0.06_{ext})\times 10^{-7}$. Model dependent fits
to the $z$-spectrum have been performed (Fig.~\ref{fig:pieeg}
(right)), obtaining the corresponding form factors and BR. The
preliminary average BR in the full kinematic range
is: \mbox{$BR = (3.08 \pm 0.04_{stat} \pm 0.08_{ext} \pm
0.07_{model})\times 10^{-7}$}. Comparison of results with previous
experiments and theoretical predictions can be found in~\cite{Ev}.
%

\end{document}